\documentclass[11pt,a4paper]{article}
\date{}
\usepackage{amsmath,amsthm}
\usepackage{graphicx}

\newtheorem*{t1}{Theorem 1}
\newtheorem*{t2}{Theorem 2}
\newtheorem*{t3}{Theorem 3}
\newtheorem*{t4}{Lemma 1}
\newtheorem*{t5}{Lemma 2}
\newtheorem*{t6}{Lemma 3}
\newtheorem*{t7}{Lemma 4}
\newtheorem*{t8}{Theorem 4}

\begin{document}

\markboth{Ch. Dangalchev}
{Closeness Centralities of Lollipop Graphs}

\title{Closeness Centralities of Lollipop Graphs}

\author{Chavdar Dangalchev 
\\ dangalchev@hotmail.com
\\ Institute of Mathematics and Informatics
\\  Bulgarian Academy of Sciences.}

\maketitle

\begin{abstract}
Closeness is one of the most studied characteristics of networks. 
Residual closeness is a very sensitive measure of graphs robustness. 
Additional closeness is a measure of growth potentials of networks.
In this article we calculate the closeness, vertex residual closeness,
link residual closeness, and additional closeness of lollipop graphs.


\end{abstract}

\section{Introduction}

The analysis of complex networks is an important subject in different fields, like 
math, chemistry, transportation, social network analysis etc. 
It has two major components - creation (growth) and stability (vulnerability).  The additional closeness  is the maximal closeness after adding a new link to a graph. The residual closeness is the minimal closeness of the graph, received after deleting a link or a vertex. The additional and the residual closeness describe the behavior of graphs in two different directions - building and destroying the network. In real life, when we have limited resources for maintaining for example a utility network, the comparison between the additional and the residual closeness of a network will tell us where to put our effort - maintaining the existing links or building a new one.

One of the most sensitive measures of network vulnerability is residual closeness. It is introduced by Dangalchev in [1], where is given a new definition for the closeness of a simple undirected graph:
\begin{equation}
\label{eq1}
C(G)=\sum\limits_i  \sum\limits_{j\ne i} 2^{-d(i,j)}, 
\end{equation}
($d(i,j)$ is the distance between vertices $i$ and $j$). 
Some of the advantages of the above definition are that it is convenient for creating formulae for graph operations (see formulae (5) and (6)) and it can be used for not connected graphs.

Let $r$  be a vertex of graph  $G$ and graph  $G_{r}$  be the graph, constructed by removing vertex $r$  and its links  from graph  $G$. 
The distance between vertices
$i$   and $j$  in  graph  $G_{r}$ is $d' (i,j)$.
Using distances $d'(i,j)$ instead of $d(i,j)$ in the above formula, we can calculate the closeness of graph $G_{r}$.
The vertex residual closeness $VR$ is defined in [1] as:
\[
VR(G)={\min_{r} \{ C(G_{r}) \}  }.
\]
If we remove only one link, instead of a vertex, we can define link residual closeness ($LR$). 
The robustness of networks is studied in many different fields – from social network analysis to  math and chemistry. Residual closeness is a strong indicator of central nodes in networks and important links in systems. The link residual closeness shows the links, critical for the existence  of transportation or utility networks. The vertex residual closeness is more important for social network analysis – it highlights the most central members of the social network.

Let $p$ and $q$  be two not connected vertices of graph  $G$ and graph  $G_{p,q}$  be the graph, constructed by adding  link $(p,q)$ to graph  $G$. 
The distance between vertices
$i$   and $j$  in  graph  $G_{p,q}$ is $d" (i,j)$.
Using distances $d"(i,j)$ instead of $d(i,j)$ in formula (1), we can calculate the closeness of graph $G_{p,q}$.
The additional closeness $A$ of graph $G$ is defined in [2] as:
\[
A(G)={\max_{p,q} \{ C(G_{p,q}) \}  }.
\]

 Calculating the additional closeness is much easier (by computational complexity)  than calculating the residual closeness - calculating the residual closeness of larger graphs is a difficult computational problem (see [3]). On the other hand, when creating formulae for closeness the situation is reversed. It is not easy to find formulae for the additional closeness even for simple graphs like path or cycle graphs (see [4]). It is important to calculate closeness for simple graphs because the closeness of more complex networks can be calculated, using the closeness of its simple parts (e.g., formulae (5) and (6)). 
The Lollipop graph is an example of a simple graph for which we can give formulae for the closeness centralities.

Lollipop graph $L_{m,n}$ has $m+n$ vertices and 
is created from complete graph $K_m$ with $m \ge 3$ vertices 
, linked by a bridge with path graph $P_n$ with
$n$ vertices. 
In [5] Barman calculates the closeness and normalized closeness of lollipop graphs, using the definitions of Freeman [6].

In sections 3,4,5, and 6 we will calculate the closeness, the vertex residual closeness, the link residual closeness, and the additional closeness of lollipop graphs. 

\section{Previous results}

In [1] is proven that
the closeness of complete graph $K_m$ is: 
\begin{equation}
\label{eq2}
C(K_m ) =\frac {m(m-1)}{2}
\end{equation}
 and the closeness of path graph $P_n$ is:
\begin{equation}
\label{eq3}
C(P_n ) = 2n - 4 + 2^{2-n}. 
\end{equation}

The closeness of the cycle graphs, proven in [1] and [7] is: 
\begin{equation}
\label{eq4}
C(C_{k+1}) = \left \{
  \begin{array}{lr}
     4p(1-3 \cdot 2^{-p-1} ),   \quad when \  k = 2p-1,\\
     2(2p+1)(1-2^{-p} ) ,     \quad when \ k = 2p.
  \end{array}
\right .
\end{equation}

Let vertex $p$ from graph $G_1$ (with vertex closeness $C(p)$ within graph $G_1$)  and vertex $q$ from graph $G_2$  (with closeness $C(q)$)  
be connected with link $(p,q)$, creating graph $G_3$.
The formula for closeness of graph $G_3$ is given in [1]:
\begin{equation}
\label{eq5}
C(G_3 )=C(G_{1} )+C(G_{2} )+\left( {1+C(p)} \right)\left( {1+C(q)} \right),
\end{equation}

Let vertex $p$ from graph $G_1$  (with vertex closeness $C(p)$) and vertex 
$q$   from graph $G_2$ (with closeness $C(q)$)  be 
collapsed into a single vertex, creating graph $G_4$.
The formula for closeness of graph $G_4$ is given in [2]:
\begin{equation}
\label{eq6}
C (G_4) =  C(G_1) + C(G_2) + 2C(p)C(q) .
\end{equation}

More information on closeness, residual closeness, and additional closeness can be found in [8-25]. Some of the articles ([7-15]) calculate the closeness and the residual closeness of certain types of graphs, others ([16],  [19], [21], [23-25]) focus on finding extremal graphs, graphs with minimal or maximal closeness, when some characteristics of graphs are fixed.

\section{Closeness of lollipop graphs}

In this section we will calculate the closeness  (by formula 1) of lollipop graphs.
Let lollipop graph $L_{m,n}$ be created from complete graph $K_m$ with $m \ge 3$ vertices 
\{$v_1$, $v_2$, ... $v_m$\}, joined by a bridge (link) with path graph $P_n$ with
$n  \ge 1$ vertices \{$v_{m+1}$, $v_{m+2}$, ... $v_{m+n}$\}.  The link connects vertices {$v_m$ and $v_{m+1}$.
We can prove:
\begin{t1}
The closeness of lollipop graph $L_{m,n}$ is:
\begin{equation}
\label{eq7}
C(L_{m,n}) = m \left( m+1- 2^{1-n}\right) 2^{-1} + 2n + 3 \cdot 2^{-n} - 3.
\end{equation}
\end{t1}
\begin{proof}The closeness of any vertex (e.g., $v_m$) of $K_m$ is  $C(v_m) = (m-1) /2$.
The closeness of end vertex (leaf) $v_{m+1}$ of $P_n$ is:
\[
C(v_{m+1}) =  2^{-1} + 2^{-2}+ ... + 2^{1-n} = 1 - 2^{1-n}.
\]
Using formula (2) for the closeness of complete graphs, 
formula (3) for the closeness of path graphs,  and 
formula (5) we receive:
\begin{align}
\begin{split}
C(L_{m,n}) &  = \frac {m(m-1)}{2} + 2n - 4 + 2^{2-n} + \left(1 + \frac {m-1}{2}\right)  
\left(1+1 - 2^{1-n}\right) 
\nonumber 
\\ &  = \frac {m(m-1)}{2} + 2n - 4 + 2^{2-n} + \frac {m+1}{2} (2 - 2^{1-n})
\\ &  = m(m+1)2^{-1} -m2^{-n} + 2n + 3 \cdot 2^{-n} - 3,
\end{split}
\end{align}
which proves the theorem.
\end{proof} 

\section{Vertex residual closeness of lollipop graphs}

In this section we will calculate the vertex residual closeness of lollipop graphs – the minimal closeness after removing a vertex and its links.
\begin{t2}
The vertex residual closeness of lollipop graph $L_{m,n}$ is:
\begin{equation}
\label{eq8}
VR(L_{m,n}) = (m-1)(m-2) 2^{-1} + 2n + 2^{2-n} - 4.
\end{equation}
\end{t2}
\begin{proof} There are 5 possible cases:

\subsection{Deleting the bridge vertex of the complete graph:} 

\noindent By deleting the bridge vertex ($v_m$) of $K_m$ we cut the link between the 
 complete graph and the path graph. The remaining graph $G_1$ has not connected
complete graph $K_{m-1}$ and path graph $P_n$. The closeness in this case is:
\[
C(G_1) = \frac {(m-1)(m-2)}{2} + 2n - 4 + 2^{2-n}.
\]

\subsection{Deleting not bridge vertex of the complete graph:} 

\noindent By deleting not bridge vertex $v_p$, $p \in [1,...,m-1]$ of $K_m$ we receive lollipop graph $G_2 = L_{m-1,n}$. The closeness of $G_2$, using formula (7), is:
\[
C(G_2) = \frac {(m-1)(m- 2^{1-n})}{2} + 2n + 3 \cdot 2^{-n} - 3.
\]
Let us compare the closenesses of graphs $G_1$ and $G_2$:
\begin{align}
\begin{split}
C(G_2) - C(G_1) &  = \frac {(m-1)(m- 2^{1-n}-m+2)}{2} +1 - 2^{-n} 
\nonumber 
\\ &  = \frac {(m-1)(2- 2^{1-n})}{2} +1 - 2^{-n} > 0,
\end{split}
\end{align}
which means that case 4.2 cannot supply the residual closeness.

\subsection{Deleting the leaf vertex of $L_{m,n}$:} 

\noindent By deleting the leaf vertex ($v_{n+m}$) of $L_{m,n}$ we receive graph $G_3 = L_{m,n-1}$. The closeness of $G_3$, using formula (7), is:
\[
C(G_3) = \frac {m \left( m+1- 2^{2-n}\right) }{2}+ 2n + 3 \cdot 2^{1-n} - 5.
\]
Comparing the closenesses of graphs $G_3$ and $G_1$ we receive:
\begin{align}
\begin{split}
C(G_3) - C(G_1) &  = \frac {m \left( m+1- 2^{2-n}\right) - (m-1)(m-2)}{2} + 3 \cdot 2^{1-n} - 1  - 2^{2-n}.
\nonumber 
\\ &  = \frac {m-m2^{2-n} +3m-2}{2} +2^{1-n} -1  
\\ &  = \frac {4(m(1-2^{-n}) -1)}{2} +2^{1-n}  >1
\end{split}
\end{align}
which means that graph $G_1$ has the minimal closeness so far.

\subsection{Deleting the bridge vertex of the path graph:} 

\noindent By deleting the bridge vertex ($v_{m+1}$) of $P_n$ we cut the link between the 
 complete graph and the path graph. The remaining graph $G_4$ has not connected
complete graph $K_{m}$ and path graph $P_{n-1}$. The closeness of $G_4$ is:
\[
C(G_4) = \frac {m(m-1)}{2} + 2n - 6 + 2^{3-n}.
\]
Comparing the closeness of graph $G_4$ with the closeness of graph $G_1$:
\begin{align}
\begin{split}
C(G_4) - C(G_1) &  = \frac {m(m-1) - (m-1)(m-2)}{2} - 2 + 2^{3-n} - 2^{2-n}
\nonumber 
\\ &  =  \frac {2(m-3)}{2} + 2^{2-n}  >1.
\end{split}
\end{align}
This means that graph $G_1$ is better than graph $G_4$.

 \quad 

\subsection{Deleting middle vertex of the path graph:} 

\noindent Let us delete a middle vertex of $P_n$, e.g. $v_k$, where $m+1<k<m+n$. 
The resulting graph $G_5$ has not connected
lollipop graph $L_{m,k-1}$ and path graph $P_{n-k}$. The closeness of $G_5$, using formula (7), is:
\begin{align}
\begin{split}
C(G_5)  &  = \frac { m ( m+1- 2^{2-k})}{2} + 2(k-1) + 3 \cdot 2^{1-k} - 3
+ 2(n-k) - 4 + 2^{2+k-n}
\nonumber 
\\ &  =  \frac { m ( m+1- 2^{2-k})}{2} + 2n - 9 + 3 \cdot 2^{1-k}  + 2^{2+k-n}
\end{split}
\end{align}
Comparing the closeness of graph $G_5$ with the closeness of $G_4$ we receive:
\begin{align}
\begin{split}
C(G_5) - C(G_4) &  =   \frac { m ( m+1- 2^{2-k})}{2} + 2n - 9 + 
3 \cdot 2^{1-k}   + 2^{2+k-n}
\nonumber 
\\ &   \quad  - \frac {m(m-1)}{2} - 2n + 6 - 2^{3-n}.
\\ &  = \frac { 2m- m2^{2-k}}{2} -3 + 3 \cdot 2^{1-k} + 2^{2+k-n} - 2^{3-n}
\\ &  = m(1- 2^{1-k})  - 3 (1 - 2^{1-k}) +  2^{2-n} (2^{k}-2)
\\ &  = (m-3)(1- 2^{1-k}) +  2^{2-n} (2^{k}-2) > 0,
\end{split}
\end{align}
which means that case 4.4 is better than case 4.5.

From all 5 cases, case 4.1 has the minimal closeness, which means that 
deleting the bridge vertex $v_m$ of the complete graph gives the vertex residual closeness of the lollipop graph
and this proves the theorem.
\end{proof}

\section{Link residual closeness of lollipop graphs}

In this section we will calculate the link residual closeness of lollipop graphs – the minimal closeness after removing a link.
\begin{t3}
The link residual closeness of lollipop graph $L_{m,n}$ is:
\begin{equation}
\label{eq9}
LR(L_{m,n}) = m(m-1) 2^{-1} + 2n + 2^{2-n} - 4.
\end{equation}
\end{t3}
\begin{proof} There are 5 possible cases:

\subsection{Deleting the bridge link $(v_{m}, v_{m+1})$:} 

\noindent By deleting the bridge link, we cut the link between the 
 complete graph and the path graph. The remaining graph $G_1$ has not connected
complete graph $K_{m}$ and path graph $P_n$. The closeness of  $G_1$ is:
\[
C(G_1) = \frac {m(m-1)}{2} + 2n - 4 + 2^{2-n}.
\]

\subsection{Deleting the leaf link $(v_{m+n-1}, v_{m+n})$:} 

\noindent By deleting the leaf link, we receive graph $G_2=L_{m,n-1}$.
The closeness in this case is:
\[
C(G_2) = \frac {m(m+1- 2^{2-n})}{2} + 2n - 5 + 3 \cdot 2^{1-n} 
\]
Let us compare the closenesses of the two cases:
\begin{align}
\begin{split}
C(G_2) - C(G_1) &  =  \frac {2m- m2^{2-n}}{2} - 1 + 3 \cdot 2^{1-n} -2^{2-n}
\nonumber 
\\ &  =  m(1-2^{1-n}) - 1 + 2^{1-n}= (m-1)(1-2^{1-n}) > 0,
\end{split}
\end{align}
which means that graph $G_2$ cannot supply the residual closeness.

\subsection{Deleting middle link $(v_{m+k}, v_{m+k+1})$ of the path graph:} 

\noindent By deleting middle link ($0<k<n-1$) we receive graph $G_3$, which has 
disconnected lollipop graph $L_{m,k}$ and path graph $P_{n-k}$.
The closeness of $G_3$  is:
\begin{align}
\begin{split}
C(G_3) &  =  m \left( m+1- 2^{1-k}\right) 2^{-1} + 2k + 3 \cdot 2^{-k} - 3
\nonumber 
\\ &   \quad  + 2(n-k) - 4 + 2^{2+k-n}
\\ &  =  m \left( m+1- 2^{1-k}\right) 2^{-1} + 2n -7 + 3 \cdot 2^{-k} + 2^{2+k-n}.
\end{split}
\end{align}
Comparing the closenesses of graphs $G_3$ and $G_1$ we receive:
\begin{align}
\begin{split}
C(G_3) - C(G_1) &  = m \left( m+1- 2^{1-k}\right) 2^{-1} + 2n -7 
+ 3 \cdot 2^{-k} + 2^{2+k-n}
\nonumber 
\\ &  \quad -  m(m-1)2^{-1} - 2n + 4 - 2^{2-n}
\\ & = m -  m2^{-k} -3 + 3 \cdot 2^{-k} + 2^{2+k-n} - 2^{2-n}
\\ & = (m - 3)(1 - 2^{-k})  + 2^{2-n}(2^{k}-1) > 0,
\end{split}
\end{align}
which means that  graph $G_3$ cannot supply the residual closeness.

\subsection{Deleting a link of the complete graph including $v_m$:} 

\noindent By deleting link $(v_m, v_p)$ we create graph  $G_4$. Only the distances
starting (ending) with vertex $v_p$ are changed:

\quad \quad - The distance $d(v_m, v_p)$ is changed from 1 to 2. The change in 
closeness is: $\Delta_1 = 2 \cdot (2^{-1} - 2^{-2} ) = 0.5$

\quad \quad - The distance between vertex $v_p$ and any vertex of the path graph is increased by 1. The change in closeness is:
\begin{align}
\begin{split}
\Delta_2 &  = 2 \left(2^{-2} + 2^{-3}+...+2^{-n-1} \right) - 2 \left(2^{-3} + 2^{-4}+...+2^{-n-2} \right)
\nonumber 
\\ &  = 2 \left(2^{-1} - 2^{-n-1} \right) - 2 \left(2^{-2} - 2^{-n-2} \right) = 0.5 - 2^{-n-1}.
\end{split}
\end{align}
The closeness of resulting graph  $G_4$ is:
\[
C(G_4) = C(L_{m,n}) - 0.5 - 0.5 + 2^{-n-1} = C(L_{m,n}) - 1 + 2^{-n-1}.
\]
The closeness of  $G_1$, following the proof of Theorem 1, could be represented as:
\begin{align}
\begin{split}
C(G_1)  &  = C(L_{m,n}) - \left(1 + \frac {m-1}{2}\right)  
\left(1+1 - 2^{1-n}\right)
\nonumber 
\\ &  = C(L_{m,n}) -(m+1) (1 - 2^{-n}) < C(G_4).
\end{split}
\end{align}
We conclude that graph  $G_1$ is better than graph  $G_4$.

\subsection{Deleting link  of the complete graph without $v_m$:} 

\noindent By deleting link $(v_p, v_q)$,  where  $p \neq m$ and  $q \neq m$, only the distance $d(v_p, v_q)$ is changed 
from $1$ to $2$. The closeness of resulting graph  $G_5$ is:
\[
C(G_5) = C(L_{m,n}) - 0.5 > C(G_1).
\]

From all 5 cases, case 5.1 has the minimal closeness, which means that 
deleting the bridge link  $(v_m, v_{m+1})$ gives the link residual closeness
of the lollipop  graph and this proves the theorem.
\end{proof} 

\section{Additional closeness - cases}

In this section we will calculate the additional closeness of lollipop graphs – the maximal closeness after adding a link.
To find the additional closeness of lollipop graphs we have to consider 3 cases:
connecting the bridge vertex $v_m$ to a vertex from the path; 
connecting not bridge vertex ($v_p$, $p = 1,2,...,m-1$) to a vertex from the path; connecting two path vertices:

 \quad - Adding link $(v_m, v_{m+k})$, where $2 \le k \le n$.

 \quad - Adding link $(v_{m-1}, v_{m+k})$, where $1 \le k \le n$. 

\quad - Adding link $(v_{m+q}, v_{m+k})$, where $1 \le q \le n-2$
and $q+2 \le k \le n$.

\noindent The second case creates a graph containing complete, 
cycle, and path graphs.
One subcase of it is when the path graph is missing: $k = n$. 
This subcase will help us to easily calculate the closeness of the second case - thus we will consider the following four cases:

 \quad - \textbf {Case A}: Adding link $(v_m, v_{m+k})$, where $2 \le k \le n$.

\quad - \textbf {Case B}: Adding link $(v_{m-1}, v_{m+k})$, where $k = n$.

\quad - \textbf {Case C}: Adding link $(v_{m-1}, v_{m+k})$, where $1 \le k \le n$.

\quad - \textbf {Case D}: Adding link $(v_{m+q}, v_{m+k})$, where $1 \le q \le n-2$

\quad \quad \quad \quad \quad \quad  and $q+2 \le k \le n$.

\noindent In the following sections we will consider each one of these cases in detail.

\subsection{Case A: Adding link $(v_m, v_{m+k})$}

Let us connect vertex $v_m$ with vertex $v_{m+k}$, $2 \le k \le n$. The resulting graph $A$
contains complete graph $K_m$, collapsed with cycle graph $C_{k+1}$ at vertex $v_m$ and
collapsed with path graph $P_{n+1-k}$ at vertex $v_{m+k}$. We can prove now:
\begin{t4}
The closeness of  graph $A$ is:
\begin{align}
\begin{split}
\label{eq10}
C(A) &  = (m+1 -  2^{k-n}) \frac {m-1}{2} +  2(n-k) - 2 + 2^{1+k-n}
\\ &  \quad +  (k+m+ 2 -  2^{1+k-n})  \frac {C(C_{k+1})}{k+1},
\end{split}
\end{align}
\[
where  \quad  \frac {C(C_{k+1})}{k+1}  = \left \{
  \begin{array}{lr}
     2-3 \cdot 2^{-p},   \quad when \   k = 2p-1,\\
     2-2^{1-p} ,     \quad when \  k = 2p.
  \end{array}
\right .
\]
\end{t4}
\begin{proof}The closeness of graph $A$, following fornula (6) is:
\[
C(A) = C(K_m) + C(C_{k+1}) + 2C(v'_m)C(v_m") + C(P_{n-k}) + 2C(v'_k)C(v_k"),
\]
where $C(v'_m)$ is the closeness of vertex $v_m$ within complete graph $K_m$;
$C(v_m")$ is the closeness of vertex $v_m$ within cycle graph $C_{k+1}$;
$C(v'_k)$ is the closeness of vertex $v_{m+k}$ within complete graph $K_m$ collapsed 
with cycle  $C_{k+1}$;
$C(v_k")$ is the closeness of vertex $v_{m+k}$ within path $P_{n-k}$.

The closeness of the complete graph is: $C(K_m) = m(m-1) /2$.
The closeness of vertex $v_m$  within the complete graph is $C(v'_m) = (m-1) /2$.
The closeness of vertex $v_m$ within the cycle graph is: 
\[
 C(v_m") = \frac{C(C_{k+1})}{k+1}.
\]
The closeness of vertex $v_{m+k}$ within the complete graph, collapsed with the cycle graph is:  
\[
 C(v'_k) = \frac{C(C_{k+1})}{k+1} + \frac {1}{2} \cdot \frac {m-1}{2}.
\]
The closeness of the path graph is: $C(P_{n+1-k}) = 2(n+1-k) - 4 + 2^{2+k-n-1}$.
The closeness of vertex $v_{m+k}$ within path graph 
$P_{n+1-k}$ is:
\[
C(v_k") =  2^{-1} + 2^{-2} +...+ 2^{-n+k} = 1 -  2^{k-n}.
\]
The closeness of graph $A$ can be calculated:
\begin{align}
\begin{split}
C(A) &  = \frac {m(m-1)}{2} + C(C_{k+1}) + 2  \frac {m-1}{2} \cdot \frac {C(C_{k+1})}{k+1}
\nonumber 
\\ &  \quad + 2(n-k) - 2 + 2^{1+k-n} + 
2 \left( \frac{C(C_{k+1})}{k+1} +  \frac {m-1}{4} \right) (1 -  2^{k-n})
\\ &  = \frac {m(m-1)}{2} + (k+m) \frac {C(C_{k+1})}{k+1} + 2(n-k) - 2 + 2^{1+k-n}
\\ &  \quad + (2 -  2^{1+k-n}) \frac {C(C_{k+1})}{k+1} + \frac {m-1}{2} (1 -  2^{k-n})
\\ &  = (m+1 -  2^{k-n}) \frac {m-1}{2} +  (k+m+ 2 -  2^{1+k-n}) \frac {C(C_{k+1})}{k+1}
\nonumber 
\\ &  \quad + 2(n-k) - 2 + 2^{1+k-n}.
\end{split}
\end{align}
From formula (4) we receive:
\[
\frac {C(C_{k+1})} {k+1} = \left \{
  \begin{array}{lr}
     2-3 \cdot 2^{-p},   \quad when \quad  k = 2p-1,\\
     2-2^{1-p} ,     \quad when \quad k = 2p,
  \end{array}
\right .
\]
which proves the Lemma.
\end{proof}

\subsection{Case B: Adding link $(v_{m-1}, v_{m+n})$}

Let us have complete graph $K_m$ with vertices \{$v_1$,...,  $v_m$\} and
path graph $P_{k}$ with vertices \{$v_{m+1}$,...,  $v_{m+k}$\}.
Let us add two more links: $(v_{m},  v_{m+1})$ (to receive lollipop graph $L_{m,k}$) 
and $(v_{m-1},  v_{m+k})$.
The resulting graph $B$ has complete graph $K_m$ joined at two points
to cycle $C_{k+2}$. 
We can prove now:
\begin{t5}
The closeness of  graph $B$ is:
\[
C(B) = \left \{
  \begin{array}{lr}
     m(m+3)2^{-1} +4p - 1 - (3p+ 2m-1)2^{-p},   \  when \   k = 2p,\\
     m(m+3)2^{-1} +4p +1 - (4p+ 3m)2^{-p-1} ,     \  when \  k = 2p+1.
  \end{array}
\right .
\]
\end{t5}
\begin{proof}
\begin{align}
\begin{split}\label{}
C (B) & =    
\sum\limits_{i=1}^{k+m} {\sum\limits_{j=1,j \ne i}^{k+m} {2^{-d(i,j)}} } =
\sum\limits_{i=1}^{m} {\sum\limits_{j=1,j \ne i}^{k+m} {2^{-d(i,j)}} } +
\sum\limits_{i=m+1}^{k+m} {\sum\limits_{j=1,j \ne i}^{k+m} {2^{-d(i,j)}} }
\nonumber 
 \\  &    =
\sum\limits_{i=1}^{m} {\sum\limits_{j=1,j \ne i}^{m} {2^{-d(i,j)}} } + 
2 \sum\limits_{i=1}^{m} {\sum\limits_{j=m+1}^{k+m} {2^{-d(i,j)}}} +
\sum\limits_{i=m+1}^{k+m} {\sum\limits_{j=m+1,j \ne i}^{k+m} {2^{-d(i,j)}}}
 \\  &    = C(K_m) 
+2 \sum\limits_{i=1}^{m-2} {\sum\limits_{j=m+1}^{k+m} {2^{-d(i,j)}}} 
+ C(C_{k+2})
- 2 \cdot 2^{-d(m-1,m)}
 \\  &    = C(K_m) + C(C_{k+2}) -1
+2 \sum\limits_{i=1}^{m-2} {\sum\limits_{j=m+1}^{k+m} {2^{-d(i,j)}}}.
\end{split}
\end{align}

If $k=2p$ then from every vertex $v_i$ ($i=1,...m-1$) of the complete graph there are 2 paths $P_p$ (which are the 2 halves of cycle $C_{k+2}$) on distance 1.
The closeness from point $v_i$ is:
\[
\sum\limits_{j=m+1}^{k+m} {2^{-d(i,j)}} = 2 \cdot 2^{-1} (2^{-1} + 2^{-2}+...+2^{-p})
=1 - 2^{-p}.
\]
The closeness of the cycle, formula (4), is:
\[
C(C_{k+2}) = C(C_{2p+2})=4(p+1)(1-3 \cdot 2^{-p-2}).
\]
Finally, the closeness of graph $B$ is:
\begin{align}
\begin{split}\label{}
C (B) & =    m(m-1)2^{-1} + 4(p+1)(1-3 \cdot 2^{-p-2}) -1 + 2(m-2)(1 - 2^{-p})
\nonumber 
\\ & =     m(m-1)2^{-1} -1 + 2m  - 4 + 4p+4- 2^{-p} (3p+3 + 2m-4)
\\  &  =    m(m+3)2^{-1} +4p -1 - (3p+ 2m-1)2^{-p}.
\end{split}
\end{align}

For the next section we need 
the  closeness of vertex $v_{m+k}$ within graph $B$. It is $1 / (k+2)$ of the  closeness of
cycle  $C_{k+2}$ plus the closeness to   $m-2$ vertices at distance 2:
\[
C(v_{m+k}) =  \frac {C(C_{k+2})}{k+2} + (m-2) 2^{-2} = 2(1-3 \cdot 2^{-p-2}) + (m-2) 2^{-2}
\]
\begin{equation}
\label{eq2}
C(v_{m+k}) = 1.5 + m2^{-2} -3 \cdot 2^{-p-1} 
\end{equation}

If $k=2p+1$ then from every vertex $i$ ($i=1,...m-1$) of the complete graph there are 2 paths $P_p$ on distance 1 and vertex $v_{m+p+1}$ on distance $p+2$.
The closeness from point $i$ is:
\[
\sum\limits_{j=m+1}^{k+m} {2^{-d(i,j)}} = 2 \cdot 2^{-1} (2^{-1} + 2^{-2}+...+2^{-p}) + 2^{-p-2}
=1 - 3 \cdot  2^{-p-2}.
\]
The closeness of the cycle, using formula (4), is:
\[
C(C_{k+2}) = C(C_{2p+3})=2(2p+3)(1-2^{-p-1}).
\]
Finally, the closeness of graph $B$ is:
\begin{align}
\begin{split}\label{}
C (B) & =    m(m-1)2^{-1} + (4p+6)(1- 2^{-p-1}) -1 + 2(m-2)(1 - 3 \cdot  2^{-p-2})
\nonumber 
\\ & =     m(m-1)2^{-1} + 2m -5 + 4p+6- 2^{-p-1} (4p+6 + 3m-6)
\\  &  =    m(m+3)2^{-1} +4p +1 - (4p+ 3m)2^{-p-1}.
\end{split}
\end{align}
The  closeness of vertex $v_{m+k}$ within graph $B$ is again $1 / (k+2)$ of the  closeness of cycle  $C_{k+2}$ plus the closeness to   $m-2$ vertices at distance 2:
\[
C(v_{m+k}) =  \frac {C(C_{k+2})}{k+2} + (m-2) 2^{-2} = 2(1- 2^{-p-1}) + (m-2) 2^{-2}
\]
\begin{equation}
\label{eq3}
C(v_{m+k
}) = 1.5 + m2^{-2} - 2^{-p},
\end{equation}
and this finishes the proof of the Lemma.
\end{proof}

\subsection{Case C: Adding link $(v_{m-1}, v_{m+k})$}

Let us have complete graph $K_m$ with vertices \{$v_1$,...,  $v_m$\} and 
path graph $P_n$ with vertices \{$v_{m+1}$,...,  $v_{m+1}$\}.
Let vertex  $v_m$ be connected  to vertex $v_{m+1}$ with link $(v_{m},  v_{m+1})$ to receive a lollipop graph. Let vertex  $v_{m-1}$ be connected to vertex $v_{m+k}$
with link $(v_{m-1},  v_{m+k})$.
The resulting graph $C$ has complete graph $K_m$, joined to cycle $C_{k+2}$ at 2 vertices
and joined to path $P_{n-k}$. We can prove now:

\begin{t6}
The closeness of  graph $C$ is:
\begin{equation}
\label{eq13}
C(C) = \left \{
  \begin{array}{lr}
      m(m+4 - 2^{-n+k}- 2^{2-p})2^{-1}  + 2n-2k  +p(4 -3 \cdot 2^{-p}) \\
     \quad \quad  \quad  - 2^{1-p}  +(3 \cdot2^{-p}-1)2^{-n+k},   \  when \  k = 2p, \\ \\
     m(m+4- 2^{-n+k} - 3 \cdot 2^{-p})2^{-1} +  2n-2k  +p(4-2^{1-p})  \\
    \quad \quad  \quad  + 2  - 2^{1-p}  + (2^{1-p}-1)2^{-n+k},    \  when \ k = 2p+1.
  \end{array}
\right .
\end{equation}
\end{t6}
\begin{proof} The closeness of vertex $v_{m+k+1}$   within  path $P_{n-k}$ is:
\[
C(v_{m+k+1}) = 2^{-1} + 2^{-2}+...+2^{-n+k+1} = 1 - 2^{-n+k+1} .
\]
Using formula (6) we receive:
\begin{align}
\begin{split}\label{}
C (C) & =   C(B) + C(P_{n-k}) +(1+C(v_{m+k})) (1+C(v_{m+k+1})) 
\nonumber 
\\ & =    C(B) + 2(n-k) - 4 + 2^{2-n+k} +(1+C(v_{m+k})) (2-2^{-n+k+1}).
\end{split}
\end{align}

If $k=2p$ we use formula (11):
\begin{align}
\begin{split}\label{}
C (C) & =   m(m+3)2^{-1} +4p -1 - (3p+ 2m-1)2^{-p} + 2(n-k) - 4 
\nonumber 
\\ &  \quad + 2^{2-n+k} + (1+1.5 + m2^{-2} -3 \cdot 2^{-p-1}) (2-2^{-n+k+1})
\\  &  =    m(m+3)2^{-1} +4p -5 - (3p+ 2m-1)2^{-p} + 2n-2k + 2^{2-n+k}
\\ &  \quad  + 5 + m2^{-1} -3 \cdot 2^{-p} - 5 \cdot 2^{-n+k} - m 2^{-n+k-1}
+ 3 \cdot2^{-n+k-p}
\\  &  =    m(m+4 - 2^{-n+k}- 2^{2-p})2^{-1}  + 2n-2k  +p(4 -3 \cdot 2^{-p}) 
\\ &  \quad  - 2^{1-p}  + 2^{2-n+k} + 3 \cdot2^{-n+k-p}- 5 \cdot 2^{-n+k}
\\  &  =    m(m+4 - 2^{-n+k}- 2^{2-p})2^{-1}  + 2n-2k  +p(4 -3 \cdot 2^{-p}) 
\\ &  \quad  - 2^{1-p}  +(3 \cdot2^{-p}-1)2^{-n+k}.
\end{split}
\end{align}

If $k=2p+1$  we use formula (12):
\begin{align}
\begin{split}\label{}
C (C) & =   m(m+3)2^{-1} +4p +1 - (4p+ 3m)2^{-p-1} + 2(n-k) - 4 
\nonumber 
\\ &  \quad + 2^{2-n+k} + (1+1.5 + m2^{-2} - 2^{-p}) (2-2^{-n+k+1})
\\  &  =    m(m+3)2^{-1} +4p - 3 - (4p+ 3m)2^{-p-1} +  2n-2k + 2^{2-n+k}
\\ &  \quad  + 5 + m2^{-1} - 2^{1-p} - 5 \cdot 2^{-n+k} - m 2^{-n+k-1}
+ 2^{-n+k-p+1}
\\  &  =    m(m+4- 2^{-n+k} - 3 \cdot 2^{-p})2^{-1} +  2n-2k  +p(4-2^{1-p})  
\\ &  \quad  + 2  - 2^{1-p}  + 2^{2-n+k}+ 2^{-n+k-p+1}- 5 \cdot 2^{-n+k}
\\  &  =    m(m+4- 2^{-n+k} - 3 \cdot 2^{-p})2^{-1} +  2n-2k  +p(4-2^{1-p})  
\\ &  \quad  + 2  - 2^{1-p}  + (2^{1-p}-1)2^{-n+k}.
\end{split}
\end{align}
This finishes the proof.
\end{proof}

\subsection{Comparison between graphs A and C}

\textbf{When $k$ is even} ($k=2p$), in appendix A is proven: 
\[
C(A) - C(C) = 1.5 - (p+2) 2^{-p} + 2^{p-n}(1- 2^{p-1})
\]

When $p =1$ then:
\[
C(A)-C(C) = 1.5 - 3 \cdot 2^{-1} = 0.
\]
In this case ($k =2$) both graphs ($A$ and $C$) could be candidates for the additional closeness.

Function $(p+2) 2^{-p}$ is decreasing with values: 1.5, 1, 0.625,...
When $p \ge 2$ we have $(p+2) 2^{-p} \le 1$. 
Using $n \ge 2p$ we receive:
\begin{align}
\begin{split}
C(A) - C(C) &  = 1.5 - (1 - 2^{1-p}) 2^{2p-n-1}  - (p+2) 2^{-p}
\nonumber 
\\ &  \ge  0.5  - (1 - 2^{1-p}) 2^{-1} = 2^{1-p} > 0.
\end{split}
\end{align}
When $p \ge 2$ graph $C$ cannot give the additional closeness.

\textbf{When $k$ is odd} ($k=2p+1$), in Appendix B is proven:
\[
C(A) - C(C) = 1.5 - (1 - 2^{1-p}) 2^{2p-n}  - (5 + 2p) 2^{-1-p}
\]
When $p =0$ then:
\[
C(A)-C(C) = 1.5 + 2^{-n} - 2.5 < 0.
\]
In this case  ($k =1$) only graph $C$ could give the additional closeness.
In fact, the link $(m,m+1)$ already exists: the added closeness 
for graph $A$ is $0$.

 When $p =1$ then:
\[
C(A)-C(C) = 1.5 -  0 - 7 \cdot 2^{-2} = -0.25.
\]
In this case  ($k =3$) graph $A$ cannot  give the additional closeness.

 When $p =2$ then $k =5$ and $n \ge 5$:
\begin{align}
\begin{split}
C(A) - C(C) &  = 1.5 - (1-0.5)2^{4-n} - 9 \cdot 2^{-3}
\nonumber 
\\ &  =  1.5  - 1.125 - 2^{3-n} \ge 0.375 - 2^{-2} = 0.125 > 0.
\end{split}
\end{align}
In this case graph $C$ cannot  give the additional closeness.

Function $(2p+5) 2^{-1-p}$ is decreasing (with values: 1.75, 1.125, 0.6875,...)
and $(2p+5) 2^{-1-p} < 1$ when $p > 2$. 
From  $n \ge k=2p +1$ we receive: $2^{2p-n} < 0.5$.
When $p \ge 3$ we have:
\begin{align}
\begin{split}
C(A) - C(C) &  = 1.5 - (1 - 2^{1-p}) 2^{2p-n}  - (5 + 2p) 2^{-1-p}
\nonumber 
\\ & >  1.5  -  (1 - 2^{1-p}) 2^{-1} - 1 = 2^{-p} > 0,
\end{split}
\end{align}
and graph $C$ cannot  give the additional closeness.

Finally, only when $k=1,2,3$ graph $C$ could  suply the additional closeness.

\subsection{Case D: Adding link $(v_{m+q}, v_{m+k})$}

Let us connect vertex $v_q$ ($2 \le q \le n-2$) with vertex $v_k$
 ($q+2\le k \le n$). The resulting graph $D$
contains lollipop graph  $L_{m,q}$, collapsed with cycle graph $C_{k-q+1}$ at vertex $v_q$ and collapsed with path graph $P_{n+1-k}$ at vertex $v_k$.

In Appendix C is calculated the closeness of graph $D$:
\begin{t7}
The closeness of  graph $D$ is:
\begin{align}
\begin{split}
C(D)  &  =  m \left( m+1- 2^{-q} -2^{k-n-q}  \right) 2^{-1} + 2(n+q-k)  - 4
\nonumber 
\\ &   \quad  +   \left( k-q+5 + (m-3)2^{-q} - 2^{1+k-n} \right)  C(C_{k-q+1})  / (k-q+1)
\\ &   \quad  +  2^{k-n} + 3 \cdot 2^{-q-1} +3 \cdot 2^{k-n-q-1}  
\end{split}
\end{align}
\end{t7}

Graph $A$ contains cycle  $C_{k+1}$ - there are two different formulae for $C(C_{k+1})$: when $k+1$ is even or odd.
Graph $D$ also contains cycle  $C_{k-q+1}$ - again we have two formulae for 
$C(C_{k-q+1})$: when $k-q+1$ is even or odd.
To compare the closenesses of graphs $A$ and $D$ we have to consider 4 cases.

In Appendix D are compared  the closenesses of graph $A$ with even number of vertices of  $C_{k+1}$ and 
graph $D$ with odd number of vertices of $C_{k-q+1}$. The result is:  $C(A) > C(D)$. 
Similarly to  Appendix D, the other 3 cases can be considered . The results of all these cases are $C(A) > C(D)$, i.e. the additional closeness cannot be $C(D)$, 
the additional closeness cannot be delivered by graph $D$.

\subsection{Finding the optimal $k$ in graph A}

When $k+1=2p$ the formula (10) for $C(A)$ is: 
\begin{align}
\begin{split}
C(A) &  = (m+1 -  2^{2p-1-n}) (m-1)2^{-1} +  2(n-2p+1) - 2 + 2^{1+2p-1-n}
\nonumber 
\\ &  \quad +  (2p-1+m+ 2 -  2^{1+2p-1-n}) ( 2-3 \cdot 2^{-p}).
\end{split}
\end{align}
The derivative $C'_p(A)$  of $C(A)$ by $p$ is:
\begin{align}
\begin{split}
C'_p(A) &  =  -  (m+3) 2^{2p-n-1} ln2 -6\cdot 2^{-p}
\nonumber 
\\ &  \quad +3  (2p+m+ 1)  2^{-p} ln2  + 3 \cdot 2^{p-n}ln2.  
\end{split}
\end{align}
If we divide $C'_p(A)$ by $3\cdot 2^{-p}ln2$ we receive:
\begin{equation}
\label{eq14}
C'_p(A) / (3 \cdot 2^{-p}ln2) = 2p + m+1 + 2^{2p-n} - (m /3+1) 2^{3p-n-1}  - 2 /ln2.  
\end{equation}

The function $C'_p(A)$ is positive  when  $p=1$ and it  continues to be positive 
until around $p \approx 0.35n$,
then it becomes negative.
Let $p^*$ be the first $p>1$ with the negative value of $C'_p(A)$. 
We compare the value $C(A(p^*))$ from formula (10) with $p=p^*$ to the value  
$C(A(p^*-1))$ from formula (10) with $p=p^*-1$ and determine the value of $k=2p-1$
supplying the bigger closeness.

Similar is the result when $k+1=2p+1$ - the formula (10) for $C(A)$ is: 
\begin{align}
\begin{split}
C(A) &  = (m+1 -  2^{2p-n}) (m-1)2^{-1} +  2(n-2p) - 2 + 2^{1+2p-n}
\nonumber 
\\ &  \quad +  (2p+m+ 2 -  2^{1+2p-n}) ( 2-2^{1-p}).
\end{split}
\end{align}
The derivative $C'_p(A)$  of $C(A)$ by $p$ is:
\begin{align}
\begin{split}
C'_p(A) &  = - (m+3)  2^{2p-n}ln2 -2^{2-p}
\nonumber 
\\ &  \quad + (2p+m+ 2)2^{1-p} ln2+  2^{2+p-n}ln2
\end{split}
\end{align}
If we divide $C'_p(A)$ by $2^{1-p} ln2$ we receive:
\begin{equation}
\label{eq15}
C'_p(A) / (2^{1-p} ln2) = 2p+m+ 2+  2^{1+2p-n}  - (m+3)  2^{3p-n-1} -2 \ ln2.
\end{equation}

The upper function is positive, when $p$ is smaller, and it
becomes negative at point $p^*$ ( $p^*$ around $0.35n$). Again, we compare the
values $C(A(p^*))$ or $C(A(p^*-1))$, using formula (10).

The additional closeness will be the one with the maximal value, after comparing
the 4 values: two with $k+1=2p$ and two with $k=2p$.

\subsection{Formulae for additional closeness of lollipop graphs}

To finish the calculation of the additional closeness $A(L_{m,n})$ of lollipop graph $L_{m,n}$ we will prove:
\begin{t8}
The additional closeness of lollipop graph $L_{m,n}$ is:
\begin{equation}
\label{eq15}
A(L_{m,n}) = \left \{
  \begin{array}{lr}
     m^2 2^{-1} + 1,   \quad when \  n = 1,\\
     m(m+1)2^{-1} + 2 ,     \quad when \  n = 2,\\
     m(2m+3)2^{-2}+ 4,     \quad when \  n = 3,\\
     m(m+2)2^{-1}+ 6,     \quad when \  n = 4,\\
     use  \  (14) , \  (15) \  and \ then  \  (10),     \quad when \ n \ge 5.  
  \end{array}
\right .
\end{equation}
\end{t8}
\begin{proof}

When $n=1$, any added link $(v_p, v_{m+1})$, $1 \le p \le m-1$ is replacing the distance $d(v_p, v_{m+1})$ from $2$ to $1$ - it is adding $0.5$ to the closeness.
We use formula (7) with $n=1$ for $C(L_{m,1})$ and receive:
\[
A(L_{m,1})  = C(L_{m,1}) + 0.5 = (m^2+1) 2^{-1} +  0.5 = m^2 2^{-1}+1.
\]
The same result we will receive if we use formula (13) with $n=1$, $k=1$, and $p=0$.
From Section 10 we know that $C(C) > C(A)$.

When $n=2$, we know from the comparison between graphs $A$ and $C$
(Section 10), that both graphs can give the additional closeness. When we add link 
$(v_m, v_{m+2})$ we have to use formula (10) with $n=2$, $k=2$, and $p=1$:
\[
A(L_{m,2})  = m \frac {m-1}{2} + (m+2) = m(m+1)2^{-1} + 2.
\]
The same result we will receive if we add link $(v_{m-1},v_{m+2})$ and we use formula (13) with $n=2$, $k=2$, and $p=1$:
\[
A(L_{m,2})  = m(m+1)2^{-1} + 2.5 - 1 + 1.5 - 1= m(m+1)2^{-1} + 2.
\]

When $n=3$, only graph $C$ can give the residual closeness. We add link 
$(v_{m-1}, v_{m+3})$  and use formula (13) with $n=3$, $k=3$, and $p=1$:
\[
A(L_{m,3})  = m(m+1.5)2^{-1} + 3 + 2 - 1 = m(2m+3)2^{-2} + 4.
\] 

When $n=4$, adding link $(v_{m-1},v_{m+3})$ will create graph $C$.
We use formula (13) with $n=4$, $k=3$, and $p=1$:
\[
C(C)  = m(m+2)2^{-1} + 8-6  +3 + 2  - 1 = m(m+2)2^{-1} + 6.
\] 
We know, from the comparison between graphs, that adding link 
$(v_m,v_{m+3})$ will create a graph with less closeness.

There is one more option: adding link $(v_m,v_{m+4})$ to receive graph $A$.
From formula (10) with $n=4$, $k=4$, and $p=2$ we receive:
\[
C(A)  = m(m-1)2^{-1} -2  +2 + (m+4) (2- 0.5)= m(m+2)2^{-1} + 6.
\] 

We can see that adding link $(v_{m-1},v_{m+3})$ or link $(v_m,v_{m+4})$
we receive the additional closeness:
\[
A(L_{m,4})  = m(m+2)2^{-1} + 6.
\] 

When $n \ge 5$ the additional closeness is supplied by adding link $(v_m,v_{m+k})$.
We use formulae (14) or (15) to determine the values of $k$ and then formula (10) to calculate the additional closeness of the lollipop graph $L_{m,n}$.
The case with two links, supplying the additional closeness, is impossible when $n \ge 5$,
because the value of $k$, given by formulae (14) or (15), is much bigger than
$3$ ($3$ is the maximal value in Section 10, giving $C(C)>C(A)$).
\end{proof}

\section{Conclusion.} 
In this article we have calculated closeness centralities of lollipop graphs $L_{m,n}$.
The formulae for closeness, link residual closeness, and vertex residual closeness are (7),(8), and (9). For smaller graphs we have calculated the additional closeness (formula (16)) and for bigger graphs we give a process for
calculating it.

\section{Appendix A. Comparing graphs A and C - even}

We will calculate $C(A)-C(C)$ where graph $A$ is created by adding link $(v_m,v_{m+k})$, graph $C$ is created by adding link $(v_{m-1},v_{m+k})$, and  $k=2p$:
\begin{align}
\begin{split}
C(A) &  = (m+1 -  2^{k-n}) \frac {m-1}{2} +  2(n-k) - 2 + 2^{1+k-n}
\nonumber 
\\ &  \quad +  (k+m+ 2 -  2^{1+k-n}) ( 2-2^{1-p}),
\end{split}
\end{align}

\begin{align}
\begin{split}
C(C) &  = m(m+4 - 2^{-n+k}- 2^{2-p})2^{-1}  + 2n-2k  +p(4 -3 \cdot 2^{-p}) 
\nonumber 
\\ &  \quad - 2^{1-p}  +(3 \cdot2^{-p}-1)2^{-n+k}.
\end{split}
\end{align}

\begin{align}
\begin{split}
C(A) - C(C) &  = (m+1 -  2^{k-n}) (m-1)2^{-1} +  2(n-k) - 2 + 2^{1+k-n}
\nonumber 
\\ &  \quad \quad +  (k+m+ 2 -  2^{1+k-n}) ( 2-2^{1-p})
\\ &  \quad \quad - m(m+4 - 2^{-n+k}- 2^{2-p})2^{-1}  - 2n+2k  
\\ &  \quad  \quad  -p(4 -3 \cdot 2^{-p}) + 2^{1-p}  - (3 \cdot2^{-p}-1)2^{-n+k}
\\ &  =m (m+1 -  2^{k-n}) 2^{-1} - (m+1 -  2^{k-n}) 2^{-1}
\\ &  \quad \quad +  2p ( 2-2^{1-p}) + (m+ 2 -  2^{1+k-n}) ( 2-2^{1-p})
\\ &  \quad \quad - m(m+4 - 2^{-n+k}- 2^{2-p})2^{-1}   - 2 + 2^{1+k-n}
\\ &  \quad  \quad -p(4 -3 \cdot 2^{-p})  + 2^{1-p}  - (3 \cdot2^{-p}-1)2^{-n+k}
\\ & = -m2^{-1} -2^{-1}  +  2^{k-n-1} + m  ( 2-2^{1-p}) 
\\ &  \quad \quad  -  p 2^{-p} + (2 -  2^{1+k-n}) ( 2-2^{1-p})
\\ &  \quad \quad - m(3 - 2^{2-p})2^{-1}   - 2 + 2^{1+k-n}
\\ &  \quad  \quad  + 2^{1-p}  - (3 \cdot2^{-p}-1)2^{-n+k}
\\ &  =   m  ( 2-2^{1-p} - 2^{-1} - 3 \cdot 2^{-1}+ 2^{1-p})
\\ &  \quad \quad  -  p 2^{-p} + 4 - 2^{2+k-n} -2^{2-p} +2^{2+k-n-p}
\\ &  \quad \quad    - 2 + 2^{1+k-n}  -2^{-1}  +  2^{k-n-1} 
\\ &  \quad  \quad  + 2^{1-p}  - 3 \cdot2^{-p-n+k}+ 2^{-n+k}
\\&  =   -p 2^{-p} + 2  -2^{-1} -2^{1-p} +2^{k-n-p}
\\ &  \quad  \quad  +  2^{k-n}(2+  2^{-1} -4 +1)
\\&  =  1.5 - (p+2) 2^{-p} +2^{k-n-p}- 2^{k-n-1}
\\&  =   1.5 - (p+2) 2^{-p} +2^{p-n}- 2^{2p-n-1}
\\&  =  1.5 - (p+2) 2^{-p} + 2^{p-n}(1- 2^{p-1})
\end{split}
\end{align}

\section{Appendix B. Comparing graphs A and C - odd}

We will calculate $C(A)-C(C)$ where graph $A$ is created by adding link $(v_m,v_{m+k})$, graph $C$ is created by adding link $(v_{m-1},v_{m+k})$, and  $k=2p+1$. Lemma 1 is proven for $k=2p-1$ - in this case we use formula (10) with 
$C(k+1) / (k+1) =2-3 \cdot 2^{-1-p}$. Formula (13) remains the same: 
\begin{align}
\begin{split}
C(A) &  = (m+1 -  2^{k-n}) \frac {m-1}{2} +  2(n-k) - 2 + 2^{1+k-n}
\nonumber 
\\ &  \quad +  (k+m+ 2 -  2^{1+k-n}) ( 2-3 \cdot 2^{-1-p}),
\end{split}
\end{align}
\begin{align}
\begin{split}\label{}
C (C) & =   m(m+4- 2^{-n+k} - 3 \cdot 2^{-p})2^{-1} +  2n-2k  +p(4-2^{1-p})  
\nonumber 
\\ &  \quad  + 2  - 2^{1-p}  + (2^{1-p}-1)2^{-n+k}.
\end{split}
\end{align}

\begin{align}
\begin{split}
C(A) - C(C) &  = (m+1 -  2^{k-n}) (m-1)2^{-1} +  2(n-k) - 2 + 2^{1+k-n}
\nonumber 
\\ &  \quad \quad +  (k+m+ 2 -  2^{1+k-n}) ( 2-3 \cdot 2^{-1-p})
\\ &  \quad \quad -  m(m+4 - 2^{-n+k}- 3 \cdot 2^{-p})2^{-1}  - 2n+2k  
\\ &  \quad  \quad  -p(4 - 2^{1-p})  -2  +2^{1-p}  - (2^{1-p}-1)2^{-n+k}
\\ &  = m(m+1 -  2^{k-n}) 2^{-1}- (m+1 -  2^{k-n}) 2^{-1}  + 2^{1+k-n}
\\ &  \quad \quad +  (k+ 2 -  2^{1+k-n}) ( 2-3 \cdot 2^{-1-p}) + m (2-3 \cdot 2^{-1-p})
\\ &  \quad \quad -  m(m+4 - 2^{-n+k}- 3 \cdot2^{-p})2^{-1}  
\\ &  \quad  \quad  -p(4 - 2^{1-p})   - 4 +2^{1-p}  - (2^{1-p}-1)2^{-n+k}
\\ &  = - 0.5 + 2^{k-n-1} + 2^{1+k-n}
\\ &  \quad \quad +  2k+ 4 -  2^{2+k-n} -3k2^{-1-p}  -3 \cdot 2^{-p}  +3 \cdot 2^{k-n-p}
\\ &  \quad \quad + m (2-3 \cdot 2^{-1-p}) -  m(4 - 3 \cdot 2^{-p})2^{-1} 
\\ &  \quad  \quad  -p(4 - 2^{1-p})   - 4 +2^{1-p}  - 2^{k-n+1-p}+2^{k-n}
\\ &  = - 0.5 + 2^{k-n} (2^{-1} + 2-4+1)
\\ &  \quad \quad +  k(2 -3\cdot 2^{-1-p}) + 2^{-p} (2 -3) 
\\ &  \quad  \quad  -p(4 - 2^{1-p})   + 2^{k-n-p}(3-2) 
\\ &  = - 0.5 + 2^{k-n} (2^{-1} -1) - 2^{-p} + 2^{k-n+1-p}
\\ &  \quad \quad +  (2p+1)(2 -3 \cdot 2^{-1-p}) -p(4 - 2^{1-p})   
\\ &  = - 0.5 - 2^{2p-n} - 2^{-p} + 2^{k-n-p} 
\\ &  \quad \quad +  4p+2 -3p2^{-p} -3 \cdot 2^{-1-p} -4p +2p2^{-p} 
\\ &  = 1.5 - 2^{2p-n} + 2^{p-n+1} -5 \cdot 2^{-1-p}   - p2^{-p}
\\ &  = 1.5 - (1 - 2^{1-p}) 2^{2p-n}  - (5 + 2p) 2^{-1-p}
\end{split}
\end{align}

\section{Appendix C. Calculating of $C(D)$}

We will calculated the closeness of graph $D$ and prove:
\begin{t7}
The closeness of  graph $D$ is:
\begin{align}
\begin{split}
C(D)  &  =  m \left( m+1- 2^{-q} -2^{k-n-q}  \right) 2^{-1} + 2(n+q-k)  - 4
\nonumber 
\\ &   \quad  +   \left( k-q+5 + (m-3)2^{-q} - 2^{1+k-n} \right)  C(C_{k-q+1})  / (k-q+1)
\\ &   \quad  +  2^{k-n} + 3 \cdot 2^{-q-1} +3 \cdot 2^{k-n-q-1}  
\end{split}
\end{align}
\end{t7}
\begin{proof}
The closeness of lollipop graph $L_{m,q}$, according to Theorem 1, is:
\[
C(L_{m,q}) = m \left( m+1- 2^{1-q}\right) 2^{-1} + 2q + 3 \cdot 2^{-q} - 3.
\]
The closeness of vertex ($v_q$) within $L_{m,q}$ is  
\[
C(v_{q}) =  2^{-1} + 2^{-2}+ ... + 2^{-q} + (m-1)2^{-q-1}= 1 + (m-3)2^{-q-1}.
\]

The closeness of cycle graph $C_{k-q+1}$ is: 
\[
C(C_{k-q+1}) = \left \{
  \begin{array}{lr}
     4p(1-3 \cdot 2^{-p-1} ),   \quad when \quad  k-q+1 = 2p,\\
     2(2p+1)(1-2^{-p} ) ,     \quad when \quad k-q+1 = 2p +1.
  \end{array}
\right .
\]
The closeness ($C(v'_{q})$)  of vertex $v_{q}$ within cycle $C_{k-q+1}$ is: 
\[
C(v'_{q}) = C(C_{k-q+1}) / (k-q+1).
\]
The closeness $C(D')$ of lollipop $L_{m,q}$ plus cycle $C_{k-q+1}$ is: 
\[
C(D') = C(L_{m,q})  + C(C_{k-q+1}) + 2C(v_{q})C(v'_{q})  
\]
The closeness of vertex ($v_k$) within lollipop $L_{m,q}$ plus cycle $C_{k-q+1}$ is: 
\[
C(v_k) = C(C_{k-q+1}) / (k-q+1) + C(v_{q}) * 0.5
\]

The closeness of path graph $P_{n+1-k}$ is: 
\[
C(P_{n+1-k}) = 2 (n+1-k) - 4 + 2^{2-n-1+k} = 2 (n-k) - 2 + 2^{1-n+k}
\]
The closeness $C(v'_k)$ of vertex $v_k$ within path $P_{n+1-k}$ is: 
\[
C(v'_k) = 2^{-1} + 2^{-2}+...+2^{k-n} = 1 - 2^{k-n}.
\]
The closeness $C(D)$ of graph $D$ is:  
\begin{align}
\begin{split}
C(D) &  =  C(D')  + C(P_{n+1-k}) + 2C(v_{k})C(v'_{k})
\nonumber
\\ &  =  C(D')  + 2 (n-k) - 2 + 2^{1-n+k} + 2C(v_{k}) (1 - 2^{k-n})
\\ & =  C(L_{m,q})  + C(C_{k-q+1}) + 2C(v_{q})C(v'_{q})  + 2 (n-k) - 2 + 2^{1-n+k}
\\ &   \quad  + 2 \left( C(C_{k-q+1}) / (k-q+1) + C(v_{q}) * 0.5  \right) (1 - 2^{k-n})
\end{split}
\end{align}

\begin{align}
\begin{split}
C(D) & =  C(L_{m,q})  + C(C_{k-q+1}) + 2 (n-k) - 2 + 2^{1-n+k}
\nonumber
\\ &   \quad  + 2 \left(1 + (m-3)2^{-q-1}\right) C(C_{k-q+1}) / (k-q+1)
\\ &   \quad  + 2 \left( C(C_{k-q+1}) / (k-q+1) + (1 + (m-3)2^{-q-1}) * 0.5  \right) (1 - 2^{k-n})
\\  &  =  m \left( m+1- 2^{1-q}\right) 2^{-1} + 2q + 3 \cdot 2^{-q} - 3 +  C(C_{k-q+1}) 
\\ &   \quad  + 2 (n-k) - 2 + 2^{1-n+k} + \left(2 + (m-3)2^{-q}\right) C(C_{k-q+1}) / (k-q+1)
\\ &   \quad  + 2 C(C_{k-q+1}) (1 - 2^{k-n}) / (k-q+1)
\\ &   \quad  +  (1 + (m-3)2^{-q-1}) (1 - 2^{k-n})
\\ &  =  m \left( m+1- 2^{1-q}\right) 2^{-1} + 2(n+q-k) + 3 \cdot 2^{-q} - 5 +  2^{1-n+k} 
\\ &   \quad  +   \left( k-q+5 + (m-3)2^{-q} - 2^{1+k-n} \right)  C(C_{k-q+1})  / (k-q+1)
\\ &   \quad  + 1 + (m-3)2^{-q-1} -  (1 + (m-3)2^{-q-1})  2^{k-n}
\\  &  =  m \left( m+1- 2^{1-q}\right) 2^{-1} + 2(n+q-k) + 3 \cdot 2^{-q} - 4 +  2^{1-n+k} 
\\ &   \quad  +   \left( k-q+5 + (m-3)2^{-q} - 2^{1+k-n} \right)  C(C_{k-q+1})  / (k-q+1)
\\ &   \quad   -  2^{k-n} + m(2^{-q-1}-2^{k-n-q-1}) -3 (2^{-q-1}-2^{k-n-q-1})
\\ &  =  m \left( m+1- 2^{1-q} + 2^{-q}-2^{k-n-q}  \right) 2^{-1} + 2(n+q-k)  - 4
\\ &   \quad  +   \left( k-q+5 + (m-3)2^{-q} - 2^{1+k-n} \right)  C(C_{k-q+1})  / (k-q+1)
\\ &   \quad    -3 \cdot 2^{-q-1}+3 \cdot 2^{k-n-q-1} + 3 \cdot 2^{-q} +  2^{k-n} 
\\ &  =  m \left( m+1- 2^{-q} -2^{k-n-q}  \right) 2^{-1} + 2(n+q-k)  - 4
\\ &   \quad  +   \left( k-q+5 + (m-3)2^{-q} - 2^{1+k-n} \right)  C(C_{k-q+1})  / (k-q+1)
\\ &   \quad  +  2^{k-n} + 3 \cdot 2^{-q-1} +3 \cdot 2^{k-n-q-1}  
\end{split}
\end{align}
This finishes the proof of the Lemma.
\end{proof}

\section{Appendix D. Comparison between closenesses of graphs A and D}

We will compare the closenesses of graph $A$ with even number of vertices of  $C_{k+1}$ and 
graph $D$ with odd number of vertices of $C_{k-q+1}$.

When $k+1=2p$ we have: 
\begin{align}
\begin{split}
C(A) &  = (m+1 -  2^{k-n}) (m-1)2^{-1} +  2(n-k) - 2 + 2^{1+k-n}
\nonumber 
\\ &  \quad +  (k+m+ 2 -  2^{1+k-n}) ( 2-3 \cdot 2^{-p}).
\end{split}
\end{align}
When $k-q+1=2r+1$ we have: 
\begin{align}
\begin{split}
C(D) &  = m(m+1 - 2^{-q}- 2^{k-n-q})2^{-1}  + 2n+ 2q-2k  -4 
\nonumber 
\\ &  \quad + 2^{k-n}  + 3 \cdot 2^{-1-q} + 3 \cdot 2^{k-n-1-q}
\\ &  \quad +  (k - q + 5 + (m-3)2^{-q} -  2^{1+k-n}) ( 2-2^{1-r}).
\end{split}
\end{align}
For the difference between 2 closenesses we have:
\begin{align}
\begin{split}
C(A) - C(D) &  = (m+1 -  2^{k-n}) (m-1)2^{-1} +  2(n-k) - 2 + 2^{1+k-n}
\nonumber 
\\ &  \quad  \quad  +  (k+m+ 2 -  2^{1+k-n}) ( 2-3 \cdot 2^{-p}).
\\ &  \quad  \quad  - m(m+1 - 2^{-q}- 2^{k-n-q})2^{-1}  - 2n- 2q+2k  +4 
\\ &  \quad  \quad  - 2^{k-n}  - 3 \cdot 2^{-1-q} - 3 \cdot 2^{k-n-1-q}
\\ &  \quad  \quad  -(k - q + 5 + (m-3)2^{-q} -  2^{1+k-n}) ( 2-2^{1-r})
\\  &  = m ( -  2^{k-n})2^{-1} - (m+1 -  2^{k-n}) 2^{-1} 
\\ &  \quad  \quad   + 2^{k-n} +  (k+m+ 2 -  2^{1+k-n}) ( 2-3 \cdot 2^{-p})
\\ &  \quad  \quad  - m( - 2^{-q}- 2^{k-n-q})2^{-1}  - 2q  +2 
\\ &  \quad  \quad   - 3 \cdot 2^{-1-q} - 3 \cdot 2^{k-n-1-q}
\\ &  \quad  \quad  -(k - q + 5 + (m-3)2^{-q} -  2^{1+k-n}) ( 2-2^{1-r})
\\ &  =   -  m2^{k-n-1}   + 1.5m - 3m2^{-p} -4.5
\\ &  \quad  \quad   - 5 \cdot 2^{k-n-1} -3 (k+ 2 -  2^{1+k-n}) 2^{-p}
\\ &  \quad  \quad  + m2^{-q-1} + m2^{k-n-q-1} -  m2^{1-q} + m2^{1-r-q}
\\ &  \quad  \quad   - 3 \cdot 2^{-1-q} - 3 \cdot 2^{k-n-1-q} - 2q +k2^{1-r}
\\ &  \quad  \quad  + ( q  + 3 \cdot 2^{-q} + 2^{1+k-n}) ( 2-2^{1-r})  + 5 \cdot 2^{1-r}
\\ &  =  m( 1.5 -  2^{k-n-1} -3 \cdot 2^{-p}  -3 \cdot 2^{-q-1} +2^{k-n-q-1}+2^{1-r-q}) 
\\ &  \quad  \quad  -4.5 - 5 \cdot 2^{k-n-1} -3k 2^{-p} -  6 \cdot 2^{-p}  + 3 \cdot 2^{1+k-n-p}
\\ &  \quad  \quad + k2^{1-r} - 2q  - 3 \cdot 2^{-1-q} - 3 \cdot 2^{k-n-1-q}  + 5 \cdot 2^{1-r}
\\ &  \quad  \quad  + 2 q  + 6 \cdot 2^{-q} + 2^{2+k-n}- q2^{1-r}  - 3 \cdot 2^{1-r-q} - 2^{2+k-n-r}
\\ &  =  m( 1.5 -  2^{k-n-1} -3 \cdot 2^{-p}  -3 \cdot 2^{-q-1} +2^{k-n-q-1}+2^{1-r-q}) 
\\ &  \quad  \quad  -4.5  -3k 2^{-p} + k2^{1-r} - q2^{1-r}
\\ &  \quad  \quad +3 \cdot 2^{k-n-1}  + 3 \cdot 2^{1+k-n-p} - 3 \cdot 2^{k-n-1-q}- 2^{2+k-n-r}
\\ &  \quad  \quad     + 5 \cdot 2^{1-r} -  6 \cdot 2^{-p}  + 9 \cdot 2^{-q-1}   - 3 \cdot 2^{1-r-q} 
\end{split}
\end{align}
Using $k = 2p-1$ and $ q = 2p-1-2r$ we receive:

\begin{align}
\begin{split}
C(A) - C(D) &  = m( 1.5 -  2^{2p-n-2} -3 \cdot 2^{-p}  -3 \cdot 2^{2r-2p} +2^{2r-n-1}+2^{2+r-2p}) 
\nonumber 
\\ &  \quad  \quad  -4.5  -3(2p-1)2^{-p} +2r2^{1-r}
\\ &  \quad  \quad +3 \cdot 2^{2p-n-2}  + 3 \cdot 2^{p-n} - 3 \cdot 2^{2r-1-n}- 2^{1+2p-n-r}
\\ &  \quad  \quad     + 5 \cdot 2^{1-r} -  6 \cdot 2^{-p}  + 9 \cdot 2^{2r-2p}   - 3 \cdot 2^{2+r-2p} 
\\ &  = m( 1.5 -  2^{2p-n-2} -3 \cdot 2^{-p}  -3 \cdot 2^{2r-2p} +2^{2r-n-1}+2^{2+r-2p}) 
\\ &  \quad  \quad  -4.5  -3p2^{1-p} +r2^{2-r}
\\ &  \quad  \quad +3 \cdot 2^{2p-n-2}  + 3 \cdot 2^{p-n} - 3 \cdot 2^{2r-1-n}- 2^{1+2p-n-r}
\\ &  \quad  \quad     + 5 \cdot 2^{1-r} -  3 \cdot 2^{-p}  + 9 \cdot 2^{2r-2p}   - 3 \cdot 2^{2+r-2p} 
\end{split}
\end{align}
We denote as $K$ the terms of $C(A) - C(D)$ which do not depend on $r$:
\begin{align}
\begin{split}
K &  = m( 1.5 -  2^{2p-n-2} -3 \cdot 2^{-p})  -4.5  -3p2^{1-p}
\nonumber 
\\ &  \quad  \quad +3 \cdot 2^{2p-n-2}  + 3 \cdot 2^{p-n} -  3 \cdot 2^{-p}  
\end{split}
\end{align}
We will consider $C(A) - C(D)$ as function $F(r)$ of $r$: 
\begin{align}
\begin{split}
F(r) &  =K + m(-3 \cdot 2^{2r-2p} +2^{2r-n-1}+2^{2+r-2p}) +r2^{2-r} + 5 \cdot 2^{1-r} 
\nonumber 
\\ &  \quad  \quad  - 3 \cdot 2^{2r-1-n}- 2^{1+2p-n-r}   + 9 \cdot 2^{2r-2p}   - 3 \cdot 2^{2+r-2p}    
\\F(r+1) &  =K + m(-3 \cdot 2^{2r+2-2p} +2^{2r-n+1}+2^{3+r-2p}) +(r+1)2^{1-r} + 5 \cdot 2^{-r} 
\\ &  \quad  \quad  - 3 \cdot 2^{2r+1-n}- 2^{2p-n-r}   + 9 \cdot 2^{2r+2-2p}   - 3 \cdot 2^{3+r-2p}   
\end{split}
\end{align}

We will prove that $F(r)$ is decreasing function:
\begin{align}
\begin{split}
F(r) - F(r+1) &  =K + m(-3 \cdot 2^{2r-2p} +2^{2r-n-1}+2^{2+r-2p}) +r2^{2-r} + 5 \cdot 2^{1-r} 
\nonumber 
\\ &  \quad    - 3 \cdot 2^{2r-1-n}- 2^{1+2p-n-r}   + 9 \cdot 2^{2r-2p}   - 3 \cdot 2^{2+r-2p}    
\\ &  \quad -K - m(-3 \cdot 2^{2r+2-2p} +2^{2r-n+1}+2^{3+r-2p}) -(r+1)2^{1-r} - 5 \cdot 2^{-r} 
\\ &  \quad    + 3 \cdot 2^{2r+1-n}+ 2^{2p-n-r}   - 9 \cdot 2^{2r+2-2p}   + 3 \cdot 2^{3+r-2p}   
\\ &  = + m(9\cdot 2^{2r-2p} - 3 \cdot 2^{2r-n-1}-2^{2+r-2p}) 
\nonumber 
\\ &  \quad  +r2^{1-r}  + 3 \cdot 2^{-r} + 9 \cdot 2^{2r-1-n}
\\ &  \quad    - 2^{2p-n-r}   - 27 \cdot 2^{2r-2p}   + 3 \cdot 2^{2+r-2p}    
\\ &  = + 9(m-3)2^{2r-2p} - 3 (m-3) 2^{2r-n-1}  -(m-3) 2^{2+r-2p}
\\ &  \quad  + 2^{-r}  (3 + 2r - 2^{2p-n})
\end{split}
\end{align}
Using  $m \ge 3$, and:
\begin{align}
\begin{split}
\Delta &  = 9 \cdot 2^{2r-2p} - 3 \cdot 2^{2r-n-1}  - 2^{2+r-2p} 
\nonumber 
\\ &  =2^{2+2r-2p} + 3 (2^{-2p}  - 2^{-n-1}) 2^{2r} + ( 2^{r+1}-4)  2^{r-2p} > 0,
\end{split}
\end{align}
and $3 + 2r - 2^{2p-n} > 0$ we receive:

\[
F(r) - F(r+1) = \Delta (m-3) + 2^{-r}  (3 + 2r - 2^{2p-n}) >0.
\]

The minimal value of $C(A)-C(D)$ is when $r$ has the maximal possible value, i.e. $r = p-1$.
\begin{align}
\begin{split}
C(A) - C(D) &  \ge m( 1.5  -  2^{2p-n-2} -3 \cdot 2^{-p}  -3 \cdot 2^{-2} + 2^{2p-n-3}+2^{1-p}) 
\nonumber 
\\ &  \quad  \quad  -4.5  -3p2^{1-p} +(p-1)2^{3-p}
\\ &  \quad  \quad +3 \cdot 2^{2p-n-2}  + 3 \cdot 2^{p-n} - 3 \cdot 2^{2p-3-n}- 2^{2+p-n}
\\ &  \quad  \quad     + 5 \cdot 2^{2-p} -  3 \cdot 2^{-p}  + 9 \cdot 2^{-2}   - 3 \cdot 2^{1-p} 
\\ &  = m( 0.75 - 2^{-p} -2^{2p-n-3}) -2.25 + 3 \cdot 2^{-p} +3 \cdot 2^{2p-n-3}
\\ &  \quad  \quad  +p2^{1-p}-2^{3-p}  - 2^{p-n} + 8 \cdot 2^{-p} 
\\ & = (m-3)( 0.75 - 2^{-p} -2^{2p-n-3}) +p2^{1-p} - 2^{p-n}    
\end{split}
\end{align}
Having $n \ge 2p-1$ we receive :
\[
p2^{1-p} - 2^{p-n} = (p-1)2^{1-p}  + 2^{1-p} (1-2^{2p-1-n}) > 0,
\]
\[
0.75 - 2^{-p} -2^{2p-n-3} =  2^{-1}- 2^{-p} +  2^{-2} (1-2^{2p-1-n})> 0.
\]
Finally, we receive  $C(A) > C(D)$, i.e. the additional closeness can be delivered only by C(A). 

If we consider the case, where $C_{k+1}$ and $C_{k-q+1}$ have odd number of vertices and the two cases where $C_{k-q+1}$ has even number of vertices, 
we will receive again $C(A) > C(D)$.

\quad

\noindent  \textbf {Data Availability}

\noindent All data are incorporated into the article.

\end{document}